\documentclass[runningheads, orivec]{llncs}

\usepackage[T1]{fontenc}

\usepackage{graphicx} 

\usepackage{hyperref}
\usepackage{color}

\usepackage{multirow}
\usepackage{booktabs}
\usepackage{xcolor}
\usepackage{pifont}
\usepackage{amsmath}
\usepackage{array}
\usepackage{bbding}

\newif\ifcomments
\commentsfalse

\ifcomments

\newcommand{\gm}[1]{{\color{blue}#1}}
\else

\newcommand{\gm}[1]{#1}
\fi

\newif\ifnotblind
\notblindtrue

\newcommand{\cost}[2]{C_{#1}\left(#2\right)}

\newcommand{\tabcolsepvalue}{8pt}
\newcommand{\arraystretchvalue}{1.2}

\newcommand{\flfsdp}{\emph{FL+FSDP}}
\newcommand{\flhsdp}{\emph{FL+HSDP}}

\newcommand{\allgather}{\texttt{all-gather}}
\newcommand{\allgathers}{\texttt{all-gather}s}
\newcommand{\reducescatter}{\texttt{reduce-scatter}}
\newcommand{\reducescatters}{\texttt{reduce-scatter}s}
\newcommand{\allreduce}{\texttt{all-reduce}}
\newcommand{\allreduces}{\texttt{all-reduce}s}
\newcommand{\broadcast}{\texttt{broadcast}}

\graphicspath{{Figures/}}

\begin{document}

\title{Accelerating Sharded Data Parallelism at Scale with Federated Learning}
\titlerunning{Accelerating Sharded Data Parallelism at Scale with Federated Learning}

\ifnotblind

\author{
Gianluca~Mittone$^{\textrm{\Envelope}}$\orcidID{0000-0002-1887-6911}
\and
Marco~Aldinucci\orcidID{0000-0001-8788-0829}
}
\authorrunning{G. Mittone, M. Aldinucci}
\institute{University of Turin, Turin, Italy\\
\email{\{gianluca.mittone,
marco.aldinucci\}@unito.it}\\
\url{https://alpha.di.unito.it}}

\fi

\begin{center}
The following paper is the accepted version of Springer copyrighted material
\\[12pt]
\textit{Gianluca Mittone, \& Marco Aldinucci (2026). Accelerating Sharded Data Parallelism at Scale with Federated Learning. In Euro-Par 2026: Parallel Processing - 32nd European Conference on Parallel and Distributed Processing, Pisa, Italy, August 24-28, 2026, Proceedings, Part II (pp. 437–451). Springer.}
\\[12pt]
presented at the EuroPar'26 conference in Pisa, Italy, and awarded with the Best Paper Award.
\\[12pt]
DOI: \href{https://doi.org/10.1007/978-3-032-35251-4_30}{https://doi.org/10.1007/978-3-032-35251-4\_30}
\end{center}

\maketitle

\begin{abstract}
\gm{The symbiotic scaling of \emph{artificial intelligence} models and \emph{high‑performance computing} systems continually creates algorithmic challenges in their convergence.
\emph{Foundation models} (FMs) are a crucial example, requiring months‑long training on thousands of cutting‑edge GPUs.
Sharded \emph{data parallelism} (DP) is the dominant strategy to accelerate such computations by splitting data and models across multiple GPUs. However, it incurs prohibitive communication overhead when deployed at scale, particularly on multi-tier interconnects with heterogeneous performance.
Inspired by the efficient communication principles of \emph{federated learning} (FL), this work introduces two hybrid algorithms—\flfsdp\ and \flhsdp—interleaving sharded DP with \emph{FedAvg}‑style aggregations.
Such approaches decouple large DP deployments into smaller, loosely‑coupled \emph{federation groups}, requiring minimal inter‑group traffic while keeping the global batch size bounded by the groups' size.
Formal analysis of communication costs and experimental validation prove their scalability and flexibility.
A Llama3.1 8B pre-training on 512 A100 GPUs shows that, under identical hyperparameters, \flfsdp\ and \flhsdp\ achieve up to 8.04$\times$ faster data processing and 4.48$\times$ lower evaluation perplexity than their counterparts, demonstrating superior computational efficiency and improved model quality.
These properties stem from reduced communication overhead and the bounded growth of the global batch size relative to the federation group size.}
\end{abstract}


\section{Introduction}
\label{sec:intro}

Contemporary large‑scale clusters and High‑Performance Computing (HPC) systems are increasingly being designed, built, and operated with \emph{artificial intelligence} (AI) workloads as their primary target.
This shift is reflected in the growing interest in low‑precision hardware~\cite{low-precision}, AI‑specific I/O patterns~\cite{io-ai-hpc}, and even network fabrics tuned for AI traffic~\cite{hummingmesh}.
At the same time, the hardware demands of AI workloads are expanding dramatically, with model sizes, training datasets, and parallelism scales all growing by orders of magnitude~\cite{compute-trends}.

\emph{Foundation Models} (FMs) exemplify this trend, with Large Language Models (LLMs) being their most widely known representatives~\cite{opt,llama}.
Training such models can saturate the world's most powerful supercomputers thanks to finely designed training pipelines that efficiently distribute computation across thousands of GPUs~\cite{fsdp,zero,megatron}.
One of the most widespread approaches to do so is \emph{data parallelism} (DP).
This technique allows multiple copies of the same model to be trained in parallel on different data batches while keeping their gradients strictly synchronised via per-batch collective communication~\cite{ddp}.
The current state of the art in this methodology is \emph{sharded} DP, which enables training even very large FMs (i.e., tens of billions of parameters) that do not fit on a single GPU.
This is achieved by "sharding" the FM parameters across many GPUs and re-collecting them when needed, thereby further increasing communication overheads linearly with the number of model replicas~\cite{fsdp}.

However, \emph{interconnect performance} often becomes the limiting factor of such an approach at scale.
While intra‑node GPU‑to‑GPU links provide $TB/s$ bandwidth (e.g., NVLink 5.0, $\approx$1.8$TB/s$), inter‑node links are typically one to two orders of magnitude slower (e.g., InfiniBand XDR, $\approx$200$GB/s$ per link).
Moreover, modern datacenters are frequently organised in hierarchical topologies—such as a two‑level fat‑tree or the Dragonfly+ architecture~\cite{dragonfly+}—where communication across different node groups incurs additional non-uniform network noise, sometimes reducing effective bandwidth by up to 50\%~\cite{de_sensi:gpu_interconnect:2024}.

\emph{Federated Learning} (FL)~\cite{mcmahan:fl:2017} offers a complementary paradigm to solve such a problem: it enables DP training between loosely coupled devices through sporadic communications, the frequency of which can be customised.
Additionally, it does not assume raw data sharing across model replicas, even though such sharing could affect model quality.
Recent literature has explored FL for large‑scale FM training, demonstrating that it mitigates communication overhead while preserving, or even improving, model quality~\cite{photon,diloco}.
However, no prior work addresses how FL can be integrated into current sharded DP techniques to enhance their performance.
Such coupling introduces an \emph{extra degree of scalability}, mitigating the constraints imposed by the underlying network infrastructure and improving computational and learning performance.

\begin{figure}[t]
    \centering
    \includegraphics[width=.9\columnwidth]{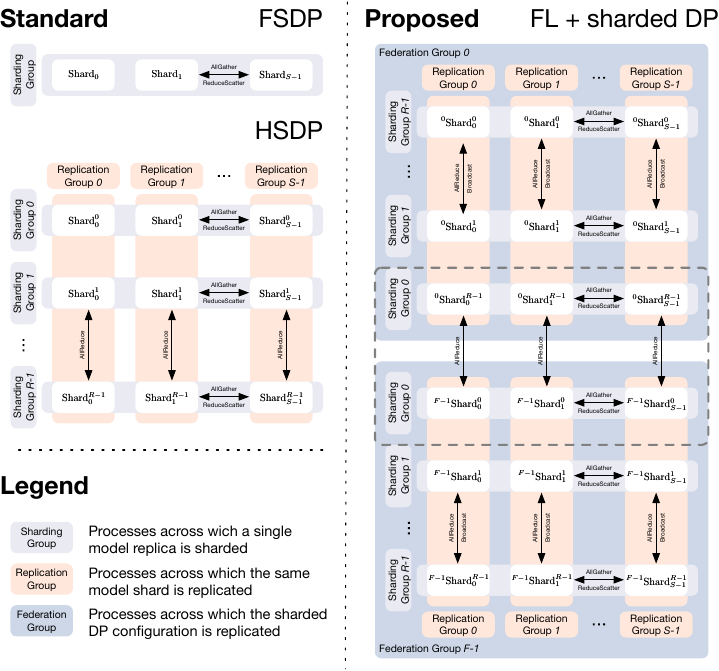}
    \caption{Standard sharded DP techniques (left) vs the proposed FL-augmented (right)}
    \label{fig:proposed}
\end{figure}

This work introduces \flfsdp\ and \flhsdp, two FL-augmented sharded DP algorithms (see Figure~\ref{fig:proposed}) improving FMs' training and learning performance at scale on modern HPC infrastructures.
The contribution is threefold:
\begin{itemize}
    \item \emph{Hybrid communication scheme}–a design that interoperates two flavours of sharded DP with FL-style aggregation, yielding higher scalability and reduced inter‑node traffic at scale, better exploiting heterogeneous interconnection performance. 
    \item \emph{Theoretical analysis}–a formal evaluation of the communication costs of fully and hybrid sharded DP and the relative FL-enriched proposed versions.
    \item \emph{Empirical validation}–extensive experiments on a top-10 Top500 HPC system demonstrating the proposed methods' scalability, communication efficiency, and convergence in a large-scale setting (128 nodes - 512 GPUs).
\end{itemize}

\section{Background}
\label{sec:background_related:background}

\subsubsection{Distributed Learning}
\label{sec:background_related:background:dp}

\gm{State-of-the-art distributed training techniques leverage several orthogonal parallelisation dimensions, such as \emph{Model} (GPipe~\cite{gpipe}), \emph{Tensor}/\emph{Sequence} (Megatron‑LM~\cite{megatron}), \emph{experts}~\cite{moe} parallelism, but \emph{data} parallelism (ZeRO~\cite{zero}) remains the most common base.
Standard DP implies training multiple replicas of the same model deployed on different computing devices simultaneously.
Each model replica is fed different data and, thanks to per-batch gradient synchronisation, the replicas' parameters evolve in perfect alignment.
This also has the collateral effect of increasing the \emph{global batch size} of training in proportion to the number of replicas, establishing a strong relation between the level of parallelism and the final model's quality.
However, DP requires that each model replica fit into each computing device's memory, which is not a valid assumption for modern FMs.
In such a case, \emph{sharded DP} extends this technique, enabling model partitioning and reconstruction across multiple devices.}


\emph{Fully Sharded Data Parallelism} (\emph{FSDP})~\cite{fsdp} (top‑left of Figure~\ref{fig:proposed}) organises the processes into a \emph{one‑dimensional} mesh of size $S$.
Each model's layer is split equally over the mesh in shards of size $M$: from this, the definition \emph{fully sharded}.
The processes sharing a sharded model replica are called \emph{sharding group}.
During the forward and backward passes, each layer must be reconstructed on each process to perform computation, implying an \allgather\ over the sharding group for each of the $L$ layers.
Similarly, during the backward pass, full gradients must be reconstructed for each layer. Thus, $L$ \reducescatters\ over the $S$ gradient shards (also of size $M$) are required to maintain replicas aligned during training.
The communication cost of FSDP is formalised in Section~\ref{sec:approach:analysis:fsdp}.

\emph{Hybrid Sharded Data Parallelism} (\emph{HSDP})~\cite{fsdp} (bottom‑left of Figure~\ref{fig:proposed}) extends FSDP by organizing the available processes into a \emph{two‑dimensional} $R\times S$ mesh: 1) $R$ \emph{sharding groups}, each constituting an independent FSDP instance, and 2) $S$ \emph{replication groups}, each connecting corresponding shards across different sharding groups.
Each sharding group keeps internal model replicas synchronised through the FSDP communication schema.
However, additional $L$ \allreduces\ within each replication group are required to synchronise corresponding gradient shards among the replication dimension.
Such \allreduces\ are mutually independent between replica groups, and can therefore be executed in parallel.
The communication cost of HSDP is formalised in Section~\ref{sec:approach:analysis:hsdp}.

\subsubsection{Federated Learning}
\label{sec:background_related:background:fl}

\gm{\emph{Federated Learning} (FL) \cite{mcmahan:fl:2017} is a widely adopted paradigm for training AI models across loosely coupled devices that have access to local, possibly private, data.
Unlike classic DP training, FL does not require frequent collective communications: synchronisation, called \emph{aggregation}, occurs only every $A$ training steps, termed \emph{rounds}, with a frequency that can range from once every few local mini‑batches to once every several local epochs, depending on the training constraints.
Such a property makes FL very adaptable to the most diverse network conditions, while allowing \emph{periodic divergence} among the model replicas.
The standard FL aggregation strategy is \emph{Federated Averaging} (\emph{FedAvg})~\cite{mcmahan:fl:2017}, an iterative algorithm which computes and redistributes back a weighted average of all locally trained models at the beginning of each federated round.}

\section{Related Works}
\label{sec:background_related:related}


Recent related work also explores FL as a means to harness loosely coupled, even geographically dispersed, compute resources for FM training.
In \emph{DiLoCo}~\cite{diloco}, Google pioneers an FL‑inspired training pipeline for FMs with 60M–400M parameters on a modest cluster (up to eight NVIDIA A100 GPUs).
The authors employ a FedOpt variant of FedAvg, using AdamW as the inner optimiser and Nesterov momentum for the outer aggregation step.
DiLoCo demonstrate faster convergence and an order‑of‑magnitude reduction in communicated data compared with a purely centralised training baseline.
On the other hand, \emph{Photon}~\cite{photon} extended this idea to a truly geo‑distributed environment.
FMs ranging from 125M to 7B parameters are trained across up to five sites on different continents (each equipped with up to two NVIDIA H100 GPUs) using a FedOpt configuration similar to DiLoCo (AdamW as inner optimiser and Nesterov's accelerated gradient as outer, i.e., FedMom~\cite{fedmom}).
Photon confirms that loosely‑coupled compute can be aggregated without sacrificing – and sometimes even improving – convergence relative to centralised training.

Both DiLoCo and Photon demonstrate that FL can be an effective alternative to conventional centralised training.
However, neither envisages FL as an \emph{intra-silo} (i.e., single-cluster) DP training strategy, nor addresses the communication patterns that arise from such a combination at scale.
The proposed contribution differs from these works in three key respects:
\begin{itemize}
    \item \emph{Intra‑silo focus}–this work targets the hierarchical networks typical of modern HPC clusters, where intra-node bandwidth is orders of magnitude higher than intra‑rack bandwidth, which is higher than inter‑rack bandwidth. 
    \item \emph{FL as a higher‑level DP primitive}–FL is employed not as a replacement for DP but as an additional, orthogonal, coarser‑grained parallelism layer that coordinates multiple DP deployments through focused communications.
    \item \emph{Scalable communication model}–the proposed three-dimensional communication topology (sharding, replication, and federation groups) reflects the physical interconnects of modern HPC systems, allowing scaling of FM training more efficiently than the performance offered by standard DP algorithms.
\end{itemize}

\section{Enhancing FSDP and HSDP Performance Through FL}
\label{sec:approach}

This research work \emph{augments sharded DP training algorithms with FL-inspired techniques} to dramatically reduce the frequency of heavyweight collectives over slow, more contended, and noisy links, while exploiting faster, more local interconnects.
This is achieved by introducing an additional, orthogonal dimension to the process topology used by sharded DP algorithms–the \emph{federation group} dimension (right side of Figure~\ref{fig:proposed}).
Training processes are first grouped into conventional sharding and replication groups, and then further partitioned into federation groups that communicate only through lightweight FL‑style aggregations.
By confining expensive collective communications to intra-federated-group levels and exploiting lightweight FedAvg‑style inter-federated-group synchronisation, the proposed algorithms can leverage the multi-tier structure of modern HPC interconnects, thereby reducing overall communication overhead.
Before delving into the proposed \flfsdp~and \flhsdp~training schemes, the communication costs of traditional state-of-the-art training techniques are presented.

\subsection{Communication Cost of SOTA Techniques}
\label{sec:approach:analysis}

\begin{table}[t]
    \caption{Table of symbols used in communication cost formulas}
    \label{tab:symbols}
    
    \centering
    \renewcommand{\arraystretch}{\arraystretchvalue}
    \setlength{\tabcolsep}{\tabcolsepvalue}
    \begin{tabular}{l p{6cm} r}
        \toprule
        \textbf{Symbol} & \textbf{Description} & \textbf{Value}\\
        \midrule

        $t$ & Time [s] & \\
        $B_n$ & Bottleneck bandwidth among $n$ processes & \\

        \midrule

        $L$ & Number of sharded model layers & $L=1$ if not sharded\\
        $S$ & Number of shards per model replica & $S=1$ if not sharded\\
        $R$ & Number of model replicas & $R=1$ if FSDP\\
        $F$ & Number of federated groups & $F=1$ if not federated\\
        $A$ & Aggregation Frequency [batches] & $A=1$ if not federated\\
        $M$ & Layer shard size [bits] & \\
        
        \midrule
        
        $\cost{AG}{D,n}$ & Cost of an \allgather among $n$ processes with message size $D$ & $t = D\frac{n-1}{nB_n}$ \\
        $\cost{RS}{D,n}$ & Cost of a \reducescatter among $n$ processes with message size $D$ & $t = D\frac{n-1}{nB_n}$ \\
        $\cost{AR}{D,n}$ & Cost of an \allreduce among $n$ processes with message size $D$ & $t = 2D\frac{n-1}{nB_n}$ \\
        $\cost{BC}{D,n}$ & Cost of a \broadcast among $n$ processes with message size $D$  & $t = \frac{D}{B_n}$ \\
        
        \bottomrule
    \end{tabular}
\end{table}

This work focuses on communication time; other overheads, such as startup time and latency, are not modelled, as their impact, although tangible, falls outside the scope of the proposed discussion.
Table~\ref{tab:symbols} summarises the used notation.
Collective communications cost formulas are based on NVIDIA NCCL~\footnote{\url{https://github.com/NVIDIA/nccl-tests/blob/master/doc/PERFORMANCE.md})}~\cite{nccl}.


\subsubsection{FSDP}
\label{sec:approach:analysis:fsdp}

As described in Section~\ref{sec:background_related:background}, FSDP and HSDP exhibit different collective communication patterns, which determine distinct cost models.
FSDP arranges the $S$ available processes into a one‑dimensional sharding group; consequently, every collective operation—whether an \allgather, \reducescatter, or \allreduce—is executed over the same mesh.
The communication cost (in terms of time) of processing one data batch through FSDP can be modelled as:

\begin{equation}
    \begin{aligned}
        C_{\text{FSDP}}
        &= 2\,\cost{AG}{M,S}L + \cost{RS}{M,S}L 
        =  ML \left( 3\frac{S-1}{SB_S} \right)
    \end{aligned}
    \label{eq:fsdp}
\end{equation}

where $B_S$ denotes the smallest bandwidth available inside the sharding group.

\subsubsection{HSDP}
\label{sec:approach:analysis:hsdp}

HSDP organises the processes into a two‑dimensional $R\times S$ mesh consisting of $R$ sharding groups and $S$ replication groups.
Each replication group acts as an independent FSDP deployment, incurring the same communication schema (two \allgathers\ and the \reducescatter\ per layer) and cost (the $3\frac{S-1}{S B_{S}}$ term of Equation~\ref{eq:fsdp}).
In addition, HSDP requires an \allreduce\ of the gradient shards across the replication groups; this operation is less frequent (once per layer during the backward pass) and can be executed in parallel over the sharding dimension.
The HSDP communication cost can thus be modelled as:

\begin{equation}
    \begin{aligned}
        C_{\text{HSDP}}
        &= 2 \cost{AG}{M,S} L + \cost{RS}{M,S} L + \cost{AR}{M,R} L \\
        &= ML\left(3\frac{S-1}{SB_S}+2\frac{R-1}{RB_R}\right) \\
    \end{aligned}
    \label{eq:hsdp}
\end{equation}

where $B_S$ and $B_R$ are, respectively, the minimal bandwidth within the sharding and replication groups.

\subsubsection{FL}
\label{sec:approach:analysis:fl}

Conversely, assuming a FedAvg aggregation strategy, FL requires calculating the weighted average of all $F$ models only every $A$ training steps.
Such an approach is usually applied in client-server architectures, but can be ported to fully decentralised topologies, as in standard DP, by arranging the $F$ processes into a one-dimensional mesh and running an \allreduce\ over the model parameters every $A$ training steps.
It can be further adapted to standard DP by running an \allreduce\ for each model layer, rather than a single one across the entire model.
Given such assumptions, the average communication cost for processing one data batch with FL through decentralised FedAvg is:

\begin{equation}
    \begin{aligned}
        C_{\text{FL}}
        &= \frac{1}{A}\cost{AR}{M,F}L 
        = ML\left[\frac{1}{A}\left(2\frac{F-1}{FB_F}\right)\right] \\
    \end{aligned}
    \label{eq:fl}
\end{equation}
where $B_F$ denotes the minimal bandwidth between the federation participants.
This work focuses on FedAvg as a base aggregation strategy, enabling easy integration of more complex ones, since most of these are built on it.

\subsection{Augmenting Sharded DP Through FL}
\label{sec:approach:proposed}

The proposed approaches share a fundamental assumption: \emph{the global processes pool is partitioned into $F$ \emph{federation groups}}, and \emph{each federation group applies the same sharding policy}. 
This guarantees one‑to‑one correspondence between shards across different federation groups, enabling correct aggregation.
Violating such an alignment constraint has severe consequences.
In the best‑case scenario, the mismatch in shard sizes would cause a runtime error, halting training.
In the worst‑case scenario, shards with different semantic values could be aggregated, leading to silently corrupted models and catastrophic learning divergence.

\subsubsection{\flfsdp}
\label{sec:approach:proposed:fsdp}

Multiple instances of the same FSDP training can be loosely synchronised through FL by a decentralised, shard-by-shard FedAvg implementation.
Each of the $F$ parallel FSDP trainings is modelled as a federation group containing a single sharding group of size $S$. 
Each federation group proceeds independently of the others, thus retaining the same $C_\text{FSDP}$ as detailed in Equation~\ref{eq:fsdp} and performing FedAvg aggregation every $A$ training steps.
Based on the decentralised FedAvg technique discussed in Section~\ref{sec:approach:analysis:fl}, such aggregation can be performed by running $S$ independent, parallel \allreduces\ for every set of $F$ corresponding shards across the federation groups for all $L$ layers.
In this way, after each aggregation, every shard is equal to the average of its corresponding shards in all the other federation groups, making the sharding groups synchronised and ready to continue training.
The average communication cost for processing one data batch with \flfsdp\ can be modelled as:

\begin{equation}
    \begin{aligned}
        C_{\text{\flfsdp}}
        = C_{\text{\emph{FSDP}}}+ \frac{1}{A}\cost{AR}{M,F}L
        = ML\left[3\frac{S-1}{SB_S}+\frac{1}{A}\left(2\frac{F-1}{FB_F}\right)\right] \\
    \end{aligned}
    \label{eq:fl_fsdp}
\end{equation}

\subsubsection{\flhsdp}
\label{sec:approach:proposed:hsdp}

As in the \flfsdp\ case, assume $F$ parallel HSDP trainings with dimensions $R\times S$, each constituting a federation group with communication cost $C_{\text{HSDP}}$ detailed in Equation~\ref{eq:hsdp}.
If $R=1$, then this scenario falls back to \flfsdp, implying the same communication cost $C_{\text{\flfsdp}}$ (Equation~\ref{eq:fl_fsdp}).
If $R>1$, then it is possible to apply the same shard-by-shard decentralised FedAvg aggregation discussed in Section~\ref{sec:approach:proposed:fsdp} across corresponding replication groups between federation groups.
Such an approach works because all replication groups within the same federation group are already synchronised with one another via HSDP, as discussed in Section~\ref{sec:background_related:background:dp}.
It is possible to save further network traffic and achieve better interconnection exploitation by running the FedAvg aggregation only between one replica per federation group and then having the updated replica \broadcast\ its parameters to the other replicas in its federation group.
In this way, only the minimal amount of information necessary is communicated between inter-federation groups, exploiting more the intra-federation-group interconnect, which is expected to be more efficient (i.e., $B_R\gg B_F$).
The average communication cost for processing one data batch with \flhsdp\ is thus:

\begin{equation}
    \begin{aligned}
        C_{\text{\flhsdp}}
        &= C_{\text{\emph{HSDP}}} + \frac{1}{A}\cost{AR}{M,F}L  + \frac{1}{A}\cost{BC}{M,R}L  \\
        &= ML\left[ 3\frac{S-1}{SB_S} + 2\frac{R-1}{RB_R}+\frac{1}{A}\left(2\frac{F-1}{FB_F}+\frac{1}{B_R}\right)\right]
    \end{aligned}
    \label{eq:fl_hsdp}
\end{equation}


\subsection{Communication Cost Comparison}
\label{sec:approach:comparison}

\gm{Table~\ref{tab:fl+dp} compares the theoretical communication cost of the base sharded DP algorithms and their FL-enhanced versions.
As can be seen, the FSDP performance is bounded by the minimum bandwidth $B_{S}$ available between any pair of processes in the sharding group, as captured by the term $3\,\frac{S-1}{S\,B_{S}}$, which dominates the communication cost.
HSDP, in turn, contributes the $2\,\frac{R-1}{R\,B_{R}}$ term to the FSDP cost, outperforming FSDP only when the intra‑sharding group network is significantly faster than the inter‑replication group one, i.e., $B_{S} \gg B_{R}$.
In practice, this situation arises when each sharding group is confined to a single physical node, benefitting from the high bandwidth of NVLink or PCIe, while replication groups span multiple nodes, relying less on InfiniBand, Ethernet, etc.
FL communication cost is always lower than FSDP, assuming the same message size and same process groups minimal bandwidth (i.e., $S=F, B_S=B_F$), and also HSDP, assuming a square mesh ($S=R=F$) and that aggregation happens on the slowest bandwidth ($B_F=B_R, B_R\ll B_S$), for any value of $A$.

Passing on the FL-enriched algorithms, it is evident that \flfsdp\ is very similar, from a communication pattern and overhead perspective, to HSDP.
HSDP \allreduces\ the sharded gradients before updating the model's parameters during each training step, while \flfsdp\ \allreduces\ the sharded model's parameters after the local update every $A$ training steps.
Models trained with HSDP thus never diverge from each other, whereas models trained with \flfsdp\ constantly diverge after each aggregation.
\flfsdp\ is advantageous over standard FSDP only if $B_S\gg B_F$, as HSDP, which always holds when the number of processes exceeds the number of GPUs available on a single compute node.
Conversely, \flhsdp\ models three tiers of interconnection performance, i.e., $B_S\gg B_R\gg B_F$, allowing better exploitation of the fastest and most efficient links while proportionally relying less on the slowest, more contended ones.
It also allows, as \flfsdp, to tune the aggregation frequency through the $A$ parameter based on the specific deployment.
These characteristics make \flhsdp\ an extremely flexible, general, and powerful sharded DP approach, opening up new possibilities in large-scale FM training.}


    

    
        
        
        
        

\begin{table}[t]
    \caption{Communication cost of FSDP, HSDP, and their FL-augmented versions}
    \label{tab:fl+dp}
    
    \centering
    \setlength{\tabcolsep}{\tabcolsepvalue}
    \renewcommand{\arraystretch}{\arraystretchvalue}
    \begin{tabular}{l c c}
        \toprule
        & \textbf{Distributed} & \textbf{\emph{Federated}} \\
        \midrule
        \textbf{FSDP} & $ML \left( 3\frac{S-1}{SB_S} \right)$ & $ML\left[3\frac{S-1}{SB_S}+\frac{1}{A}\left(2\frac{F-1}{FB_F}\right)\right]$\\
        \addlinespace[4pt]
        \textbf{HSDP} & $ML\left(3\frac{S-1}{SB_S}+2\frac{R-1}{RB_R}\right)$ & $ML\left[ 3\frac{S-1}{SB_S} + 2\frac{R-1}{RB_R}+\frac{1}{A}\left(2\frac{F-1}{FB_F}+\frac{1}{B_R}\right)\right]$ \\
        \bottomrule
    \end{tabular}
\end{table}

\section{Experimental Evaluation}
\label{sec:experiments}

\subsubsection{Experimental Setup}
\label{sec:experiments:setup}

Experimental \flfsdp\ and \flhsdp\ implementation exploiting PyTorch $v2.10.0$ and NVIDIA NCCL $v2.27.5$ are built into \emph{cross-Facility Federated Learning} (xFFL)\footnote{
\ifnotblind
\url{https://github.com/alpha-unito/xffl/tree/FL+DP}
\else
Redacted for double-blind review purposes.
\fi
}~\cite{COLONNELLI20243}, an extensive, open-source, research-oriented Python framework integrating tools for deploying large-scale FM trainings, both on single- or cross-facility scenarios.
A brief pre‑training of \emph{Llama3.1 8B}~\cite{llama} on $\approx$150M tokens from the \emph{clean\_mc4\_it} dataset~\cite{cleanmc4it} is selected as benchmark.
All runs share the same hyper-parameter setup: \texttt{bfloat16} precision, 65,536 training samples, 4,096 test samples, batch size 2, AdamW optimiser, learning rate 0.0003, aggregation every 8 steps, cosine‑decay scheduler with 10\% warm‑up steps, and same RNG configuration.

Experimental evaluation is carried out on CINECA’s \emph{Leonardo}—currently the tenth most powerful Top500 HPC~\cite{top500}—using 128 nodes.
Each node comprises a single‑socket 32‑core Intel Xeon Platinum 8358 CPU, 8$\times$64 GB DDR4 RAM, 4$\times$ custom NVIDIA Ampere A100 GPUs (64 GB), 4$\times$ NVLink 3.0 links (4$\times$200 Gbit/s per link) connecting the GPUs, and 2$\times$ dual‑port HDR100 NICs (400 Gbit/s).
Nodes are linked via an NVIDIA Mellanox DragonFly+ fabric that provides 200 Gb/s bandwidth between any pair of nodes.
A more extensive set of results—omitted here for space constraints—is publicly available on WandB\footnote{
\ifnotblind
\url{https://wandb.ai/alpha-unito/FL+DP/workspace?nw=fe1n0p5w1r7}
\else
Redacted for double-blind review purposes.
\fi
}.

\subsection{Learning Results}
\label{sec:experiments:learning}

\begin{figure}[t]
    \centering
    \includegraphics[width=\columnwidth]{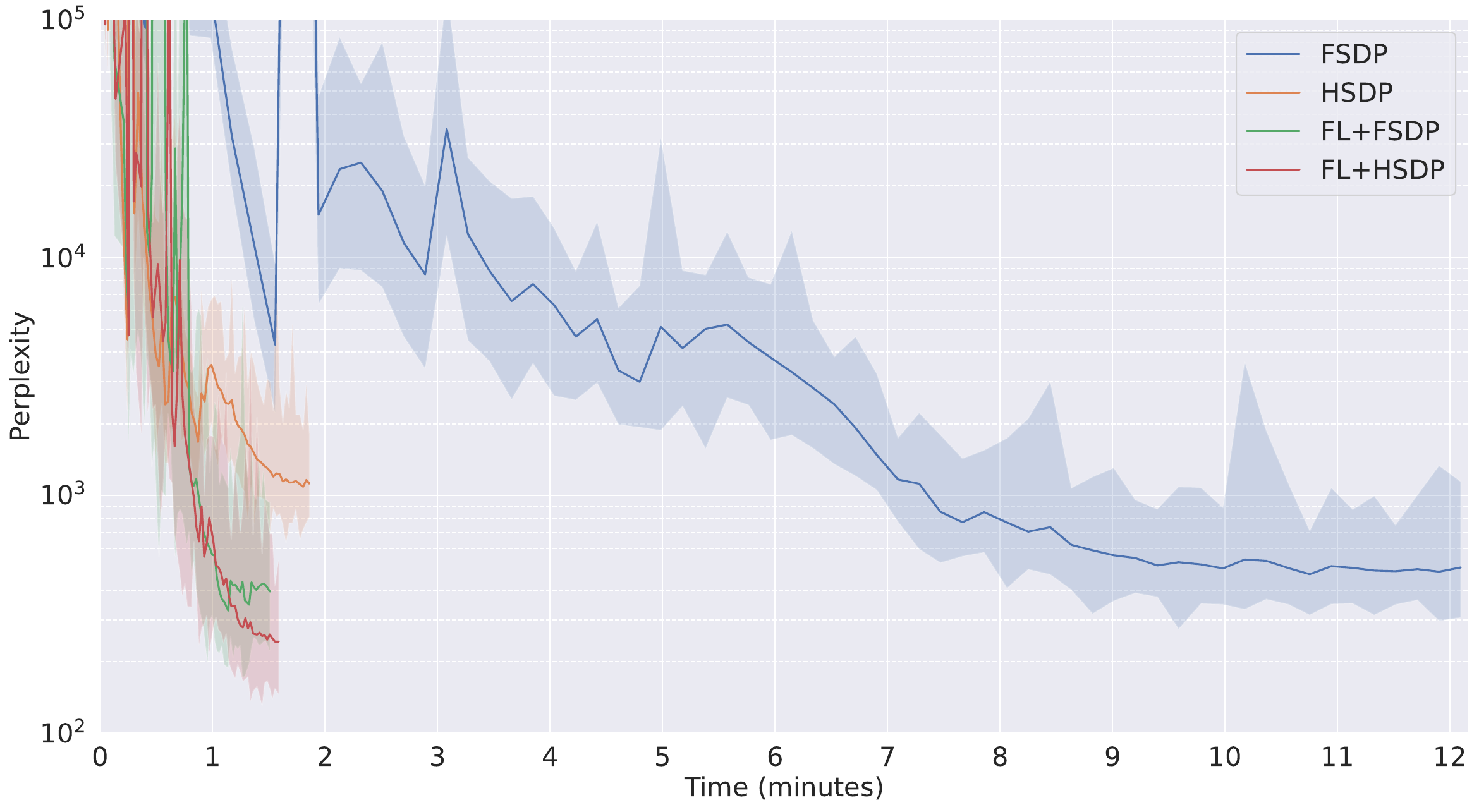}
    \caption{Training perplexity of Llama3.1‑8B on $\approx$150M tokens from the \texttt{clean\_mc4\_it} dataset, measured on 512 GPUs under a common hyper‑parameter configuration}
    \label{fig:time_to_perp}
\end{figure}

Figure~\ref{fig:time_to_perp} and Table~\ref{tab:results_overall} report the obtained perplexity for FSDP, HSDP, \flfsdp, and \flhsdp.
FSDP, assumed as a baseline, achieves 2.32$\times$ lower train perplexity than HSDP, despite being theoretically equivalent.
The different communication patterns can explain this result: HSDP averages gradients via a two‑tier \allreduce, whereas FSDP performs a single, global one.
The resulting change in floating‑point operation order, amplified by the limited precision of \texttt{bfloat16}, can lead to divergent numerical trajectories, especially for large models.
Conversely, both FL‑augmented approaches converge markedly better than their counterparts.
\flfsdp\ improves train perplexity by 1.30$\times$ over FSDP and 2.86$\times$ over HSDP.
\flhsdp\ achieves the lowest perplexity of all, obtaining better results than FSDP, HSDP and \flfsdp\ by factors of 1.95$\times$, 4.53$\times$, and 1.59$\times$, respectively.
The same is observed on the test set: \flhsdp\ outperforms FSDP, HSDP, and \flfsdp\ by 1.83$\times$, 4.48$\times$, and 1.71$\times$, confirming the learning improvement.
These gains can stem from the reduced effective global batch size implied by the FL‑augmented schemes.
By partitioning the workload into loosely‑coupled federation groups, the global batch grows only with the number of groups—not with the total number of processes—resulting in smaller, more stable batches.
Periodic FedAvg aggregations then stabilise convergence at the global level, yielding higher learning performance.

\subsection{Computational Performance}
\label{sec:experiments:performance}

\begin{table}[t]
    \caption{Llama3.1 8B train and evaluation results over $\approx$150M tokens over 128 Leonardo nodes (512 GPUs) - aggregation every 8 steps}
    \label{tab:results_overall}
    
    \centering
    \setlength{\tabcolsep}{14pt}
    \renewcommand{\arraystretch}{\arraystretchvalue}
    \begin{tabular}{l r r r r}
        \toprule
         & \multicolumn{2}{c}{\textbf{Training}} & \multicolumn{2}{c}{\textbf{Evaluation}} \\ 
         & Perplexity & Time [s] & Perplexity & Time [s] \\ 
        \midrule
        \textbf{FSDP} & 496.00 & 770.54 & 486.64 & 15.26 \\
        \textbf{HSDP} & 1151.47 & 117.37 & 1145.78 & 1.83 \\
        \textbf{\flfsdp} & 403.29 & \textbf{95.87} & 436.65 & \textbf{1.80} \\
        \textbf{\flhsdp} & \textbf{254.31} & 101.04 & \textbf{255.81} & 1.89 \\
        \bottomrule
    \end{tabular}
\end{table}

Figure~\ref{fig:time_to_perp} and Table~\ref{tab:results_overall} show that FSDP is the slowest method at this scale among those evaluated.
In contrast, HSDP, \flfsdp\ and \flhsdp\ exhibit comparable runtimes.
\flfsdp\ achieves the best throughput, processing the same amount of data 8.04$\times$, 1.22$\times$, and 1.05$\times$ faster than FSDP, HSDP, and \flhsdp, respectively.
The same holds for the evaluation phase, where \flfsdp\ is 8.48$\times$, 1.02$\times$, and 1.05$\times$ faster than FSDP, HSDP, and \flhsdp.
These performance gains can be traced back to the different communication patterns involved.
FSDP relies on a flat communication structure, incurring frequent, large‑scale collectives across many GPUs that are severely slowed by the higher‑level, noisy inter‑node links.
By contrast, HSDP and the FL‑augmented schemes employ multi‑tier communication patterns, exploiting intra‑node, inter- node, and even higher‑level connections to keep most traffic on the local, faster links, while relying proportionally less on the slower ones.

Table~\ref{tab:results_batch} breaks down the training runtime into each step sub-phase: \emph{forward}, \emph{backward}, \emph{aggregation}, and \emph{optimization}.
The \emph{forward pass} performance of FSDP is the slowest, up to 12.24$\times$ slower than the counterparts, which are instead almost aligned.
This advantage stems from the better locality of communication in these algorithms, which exploit the higher bandwidth of local interconnects.
However, the HSDP \emph{backward pass} performance is the one that suffers more, in proportion to the forward pass, mainly due to the switch to cluster-wise gradient reduction collectives.
Conversely, FL‑augmented approaches avoid such slowdowns by restricting gradient reduction to federation groups, dramatically reducing the volume of data sent over the slowest links, and are thus up to 8.43$\times$ and 1.61$\times$ faster than FSDP and HSDP.
The \emph{optimisation} step is practically 0 for FSDP due to the high level of sharding in the proposed setup; in contrast, in the other algorithm, it is constant because of the fixed-size replica group.
However, FL introduces the \emph{aggregation} step: communication is run over the slowest, busiest links in the cluster, and as such exhibits high variance.
Still, aggregation occurs only once every $A$ training steps, so its frequency and per-batch impact remain limited.
Overall, \flfsdp\ processes a mini‑batch up to 8.30$\times$ faster than FSDP, while matching HSDP’s performance (the exact ratio depends on the chosen aggregation frequency).
FL-augmented approaches thus offer up to 8.30$\times$ and 1.27$\times$ better per-batch processing performance than FSDP and HSDP, while also implying higher variances due to differences in execution time between training steps that require or do not require aggregation.

\begin{table}[t]
    \caption{Llama3.1 8B training time detail over 128 Leonardo nodes (512 GPUs) $\pm$ standard deviation (seconds) - aggregation every 8 steps}
    \label{tab:results_batch}
    
    \centering
    \setlength{\tabcolsep}{5pt}
    \renewcommand{\arraystretch}{\arraystretchvalue}
    \begin{tabular}{l r r r r r}
        \toprule
        & \textbf{Forward} & \textbf{Backward} & \textbf{Optimizer} & \textbf{Aggregation} & \textbf{Overall} \\  
        \midrule
        \textbf{FSDP}       & 4.53{\scriptsize $\pm$0.16}           & 7.50{\scriptsize $\pm$0.20}           & \textbf{0.00}{\scriptsize $\pm$0.00}  & \textbf{0.00}{\scriptsize $\pm$0.00}  & 12.04{\scriptsize $\pm$0.26} \\
        \textbf{HSDP}       & \textbf{0.37}{\scriptsize $\pm$0.00}  & 1.43{\scriptsize $\pm$0.15}           & 0.03{\scriptsize $\pm$0.00}           & \textbf{0.00}{\scriptsize $\pm$0.00}  & 1.84{\scriptsize $\pm$0.15} \\
        \textbf{\flfsdp}    & \textbf{0.37}{\scriptsize $\pm$0.00}  & \textbf{0.89}{\scriptsize $\pm$0.00}  & 0.03{\scriptsize $\pm$0.00}           & 0.17{\scriptsize $\pm$0.55}           & \textbf{1.45}{\scriptsize $\pm$0.54} \\
        \textbf{\flhsdp}    & 0.38{\scriptsize $\pm$0.02}           & 0.99{\scriptsize $\pm$0.01}           & 0.03{\scriptsize $\pm$0.00}           & 0.14{\scriptsize $\pm$0.54}           & 1.53{\scriptsize $\pm$0.54} \\
        \bottomrule
    \end{tabular}
\end{table}

\section{Discussion}
\label{sec:discussion}

The FL‑augmented sharded DP algorithms reduce processing time by up to 8.04$\times$ and improve convergence by up to 4.48$\times$ compared to their standard counterparts in the proposed scenario.
These results lead to two observations: 1) \emph{standard FSDP does not scale to hundreds or thousands of GPUs} due to its flat communication topology, forcing every collective to involve all processes and becoming a bottleneck; and 2) HSDP mitigates such problem through a two‑tier communication hierarchy, but the gain in raw throughput can be offset by \emph{noticeable degradation in learning performance}.
From a computational standpoint, the proposed FL-augmented algorithms successfully model a two-tier (\flfsdp) and a three-tier (\flhsdp) communication topology, giving fine‑grained control over the communications pattern and minimising traffic on heavily contended links.
From a learning perspective, FL-augmented approaches prevent the \emph{global batch size} from growing linearly with the number of processes. Instead, they bound it to the federation group size, yielding markedly better convergence and reducing dependence on aggressive hyper‑parameter tuning.
Finally, experimental results confirm that \flfsdp’s runtime aligns with HSDP’s, as predicted by the communication‑cost model discussed in~\ref{sec:approach:proposed:fsdp}; the aggregation frequency governs the remaining difference.

\flfsdp\ delivers superior computational performance to \flhsdp.
Although the three‑tier topology modelled by the latter would be expected to perform better, the observed advantage of \flfsdp\ can be explained by two factors: 1) the bandwidth difference between intra‑ and inter‑federation groups is insufficient to compensate for the extra \broadcast communication step introduced by \flhsdp, and 2) SLURM’s nondeterministic node allocation splits federation groups across different interconnection groups, breaking the assumed logical hierarchy.
Even so, the performance gap is small compared with the disparity in final perplexity: \flfsdp\ is only $\approx$5.39\% faster than\flhsdp, but \flhsdp\ achieves a $\approx$70.69\% improvement in perplexity.
\section{Conclusions and Future Works}
\label{sec:conclusions}

This work introduces \flfsdp\ and \flhsdp, two novel FL-augmented DP training algorithms that combine state-of-the-art sharded DP methods (FSDP, HSDP) with FL techniques (FedAvg).
By modelling a three-tier communication hierarchy—sharding, replication, and federation groups—the schemes dramatically reduce inter‑node traffic and the overall communication overhead.
Both approaches are evaluated through a formal theoretical analysis of their communication costs and large-scale experiments on a LLM pre‑training task.
The results show up to 8.04$\times$ faster data processing and 4.48$\times$ lower evaluation perplexity compared with the baseline methods.
These findings confirm that FL can be considered a “higher‑level” DP primitive, useful not only for cross‑cluster training but also for scaling state‑of‑the‑art training of FMs within a single one.

Future work will extend the methodology to other parallelism axes—MP, TP/SP, and expert EP—thereby treating FL as a fifth orthogonal dimension of FM parallelism.
Further research will also automate the selection of sharding, replication, and federation‑group sizes based on the underlying interconnect topology, eliminating manual tuning while maximising performance.


\begin{credits}
\subsubsection{Acknowledgements and Artifact Availability}
\ifnotblind
This research work is funded by the Spoke “FutureHPC \& BigData” of the ICSC - Centro Nazionale di Ricerca in “High Performance Computing, Big Data and Quantum Computing”, by the European Union - NextGenerationEU, by the EuroHPC-JU funding under grant No. 101093441, with support from the Horizon-EuroHPC-JU-2021-COE-01 (SPACE CoE) and by the DYMAN project, funded by the European Union - European Innovation Council under G.A. n. 101161930. The authors also acknowledge the computational support kindly offered by the Abdus Salam International Centre for Theoretical Physics (ICTP), Trieste, Italy.
The artifact is available in the Zenodo repository~\cite{mittone_2026_20528934}.
\else
Redacted for double-blind review purposes.
\fi

\subsubsection{\discintname}
The authors have no competing interests to declare that are relevant to the content of this article.
\end{credits}


%
\bibliographystyle{splncs04}
\bibliography{bibliography_short}
\end{document}